\documentclass[twocolumn,showpacs,amsmath,amstex,amssymb,mathfonts,prb]{revtex4-1}
\pdfoutput=1
\usepackage{amsthm,amsfonts,graphicx,verbatim,times}
\usepackage{amsmath}
\usepackage{amssymb}
\usepackage{amsthm}
\usepackage{amsfonts}
\usepackage{listings}
\usepackage{epstopdf}
\usepackage{enumerate}
\usepackage{latexsym}
\usepackage{psfrag}
\usepackage{bm}
\usepackage{graphicx}
\usepackage{subfigure}
\newcommand{\be}{\begin{equation}}
\newcommand{\ee}{\end{equation}}
\newcommand{\bea}{\begin{eqnarray}}
\newcommand{\eea}{\end{eqnarray}}

\renewcommand{\phi}{\varphi}
\renewcommand{\epsilon}{\varepsilon}

\begin{document}

\title{Quasiholes of 1/3 and 7/3 quantum Hall states: size estimates via exact diagonalization and density-matrix renormalization group}

\author{Sonika Johri$^{1}$, Z. Papi\'c$^{1,2,3}$, R. N. Bhatt$^{1}$, and P. Schmitteckert$^{4}$}
\affiliation{$^1$ Department of Electrical Engineering, Princeton University, Princeton, NJ 08544, USA}
\affiliation{$^2$ Perimeter Institute for Theoretical Physics, Waterloo, ON N2L 2Y5, Canada} 
\affiliation{$^3$ Institute for Quantum Computing, Waterloo, ON N2L 3G1, Canada}
\affiliation{$^4$ Institute of Nanotechnology, Karlsruhe Institute of Technology, 76344 Eggenstein-Leopoldshafen, Germany}


\date{\today}

\begin{abstract}
We determine the size of the elementary quasihole in $\nu=1/3$ and $\nu=7/3$ quantum Hall states
via exact-diagonalization and density-matrix renormalization group calculations on the sphere and cylinder,
using a variety of short- and long-range pinning potentials. The size of
the quasihole at filling factor $\nu=1/3$ is estimated to be $\approx 4\ell_B$, and that of $\nu=7/3$ is
$\approx 7\ell_B$, where $\ell_B$ is the magnetic length. In contrast, the size of the Laughlin quasihole, expected to capture the basic physics in these two states, is around $\approx 2.5\ell_B$. Our work supports the earlier findings that the quasihole in the first excited Landau level is significantly larger than in the lowest Landau level.  
\end{abstract}
\maketitle

\section{Introduction}

The quantum Hall state at filling factor $\nu=1/3$ is a paradigm of a strongly-correlated and topologically ordered phase with anyonic excitations. It is described by the Laughlin wave function~\cite{laughlin}, which received compelling support by exact diagonalization of small finite-size systems~\cite{rh85}. The state is experimentally realized when the lowest Landau level (LLL) of a two-dimensional electron gas (for example, in a semiconductor heterostructure) contains one electron per three magnetic flux quanta. Its main experimental signature is a quantization of the transverse conductance and simultaneous vanishing of the longitudinal resistivity in transport~\cite{tsg}. 

However, an analogous state can be envisioned in a situation when the first excited Landau level (LL1) is 1/3-filled, while the LLL for both up and down spin components is fully filled, corresponding to total filling of 7/3. Indeed, experiments detect a quantized plateau at $\nu=7/3$, although significantly weaker than in the LLL~\cite{kumar}. 

If the two filled copies of the LLL are considered inert, the state in the valence LL1 should be similar to the Laughlin state, though not equivalent because the effective Coulomb interaction, projected to LL1, differs by a form-factor from the one projected to the LLL~\cite{prange}. The change of the interaction due to form-factor can be significant: for example, at half filling of the LLL, a compressible, Fermi-liquid-like state is realized, whereas a quantized state (possibly related to the Moore-Read Pfaffian state~\cite{Moore91}) occurs at half-filling of LL1. Therefore, although it is plausible that 1/3 and 7/3 states should be similar to each other, one must resort to numerical calculations to verify such an assertion.  

In contrast to $\nu=1/3$, numerical calculations on $\nu=7/3$ have been more scarce. As will become obvious in the following, finite-size effects at 7/3 are drastically larger than at 1/3, requiring very large system sizes that are on the verge of the current limit of exact numerical diagonalization. This necessitates approximate tools such as the density-matrix renormalization group~\cite{white} (DMRG). It was noted early on by Haldane~\cite{prange} that the ground state of 7/3 is in the vicinity of a compressible phase. Small perturbations, for example by softening the Coulomb interaction as a result of finite sample-thickness, were shown to drive the system into the incompressible phase~\cite{prds}. Given the compelling accuracy of Laughlin's theory for 1/3, it is thus plausible that the phase diagram at $\nu=7/3$ indeed contains the state in the same universality class, with the caveat that there may be other compressible phases in the vicinity, which complicate the theoretical analysis. Most recently, Ref.~\onlinecite{jainqh} studied the quasiparticles in $\nu=1/3$ and $\nu=7/3$ states using ``composite-fermion diagonalization", and reporting that latter are surprisingly larger objects in size.

In light of the above, a quantitative comparison of $\nu = 1/3$ and $\nu = 7/3$ states from the perspective of the size of their elementary quasiparticle excitations using direct numerical calculations for realistic interactions seems extremely desirable. In this paper, we describe such a study, using unbiased exact diagonalization as well as density-matrix renormalization group methods. Furthermore, we discuss two types of boundary conditions -- sphere and cylinder geometry -- each of which has its unique strengths and weaknesses from the point of view of numerical calculations. Bootstrapping the results from these different techniques and boundary conditions, we establish reliable bounds for the sizes of quasiholes at $\nu=1/3$ vs. $\nu=7/3$, providing further support for the conclusion in Ref.~\onlinecite{jainqh}. 

The size of a quasihole provides a fundamental characterization of a given quantum Hall state. For model quantum Hall wave functions, quasiholes can be created by a simple flux-insertion operation, and their sizes can be estimated using Jack polynomials~\cite{jack} or analytical matrix-product state ansatz based on conformal field theory~\cite{mps}. However, quasiholes of generic (Coulomb interaction) ground states may have significant corrections from these values, and must be computed separately. In practice, this is done by diagonalizing the interaction Hamiltonian with a specially designed ``impurity" potential at a given location, whose role is to create a charge deficit and localize a quasihole at a given point~\cite{prodan, toke, morf}. In order to do this, the magnetic flux also must be increased by an appropriate number of units with respect to the ground state. Because the flux has been increased, the state is now compressible, and the size of the quasihole will in general depend on the pinning potential. If the ground state is robust (large excitation gap), various types of pinning potentials can be used to localize the quasiholes, and the resulting size is largely insensitive to the details of the potentials. However, in the case of fragile states, such as the 7/3 state, we will see below that only certain kinds of potentials are able to localize the quasihole, and require special parameter tuning.         

In addition to its size, in a system of more than one quasihole, it is possible to braid different quasiholes around one another, and compute the associated Berry phases that determine their mutual statistics. This type of calculation requires creating at least two quasiholes and ensuring they are well-separated from each other, for which it is necessary to have a reliable estimate of their size. More importantly, similar measurements can be realized in actual interferometry experiments\cite{willett, camino}, where the present results should prove useful as a guideline for the size of the quantum point contacts through which the quasiholes are expected to tunnel.

The remainder of this paper is organized as follows. In Sec.~\ref{method} we provide a brief outline of the quantum Hall problem in the spherical and cylinderical geometry, with an emphasis on different types of pinning potentials. We also briefly discuss the implementation of our methods, exact diagonalization and density-matrix renormalization group. In Sec.~\ref{sphere}, we present the results for $\nu=1/3$ and $\nu=7/3$ quasihole in the spherical geometry. In Sec.~\ref{cylinder} we present analogous results for the cylinder geometry and highlight some special aspects of this boundary condition. We conclude with a brief discussion in Sec.~\ref{conclusions}. 

\section{Method}\label{method}

In the present work we consider a finite quantum Hall system on the surface of a sphere~\cite{Haldane86} or a cylinder~\cite{hr_cylinder}. The two geometries are formally similar, but a finite cylinder has two open boundaries (the electron density can extend beyond the edges even though the number of orbitals and electrons in the finite-size system is a constant), and allows the edge excitations that are absent on the sphere. Spherical geometry is discussed at length in Ref.~\onlinecite{jainbook}, and additional details on cylinder geometry will be provided elsewhere~\cite{sonika}. 

In both cases, we consider $N$ electrons in an area enclosing $2S$ flux quanta. For the sphere, the strength of the magnetic flux automatically defines the radius of the sphere: $R=\sqrt{S}\ell_B$. In the case of cylinder, the flux $2\pi(2S+1)$ must be equal to the area of the cylinder $L\times H$, where $L$ is the perimeter, and $H$ is the height of the cylinder. This naturally introduces the aspect ratio, $a=L/H$, as an additional parameter. Fluid states are expected to be realized in the regime $L\sim H$, hence $a\approx 1$. If one of the dimensions of the cylinder is much smaller than the other, states with other types of order become preferred~\cite{sonika}.   

In order to study a single quasihole, we choose $2S=3(N-1)+1$, i.e. one unit of flux in addition to the ground state flux of the Laughlin state. In this case, the system contains a single quasihole excitation with fractional charge~\cite{laughlin} $|e^{\ast}|=e/3$. For the Laughlin wave function, $\Psi_L(z_1,..,z_N)=\prod_{i<j}(z_i-z_j)^3$, it is straightforward to construct a quasihole localized at a given point $z_0$:
\begin{equation}\label{laughlinqh}
\Psi_{qh}(z_0)=\prod_{i}(z_i-z_0)\Psi_L(z_1,..,z_N).
\end{equation}
It can be easily shown that such a defect has the correct charge $|e^{*}|=e/3$ and obeys anyonic statistics~\cite{arovas}. 

\begin{figure}[htb]
\includegraphics[width=0.4\linewidth]{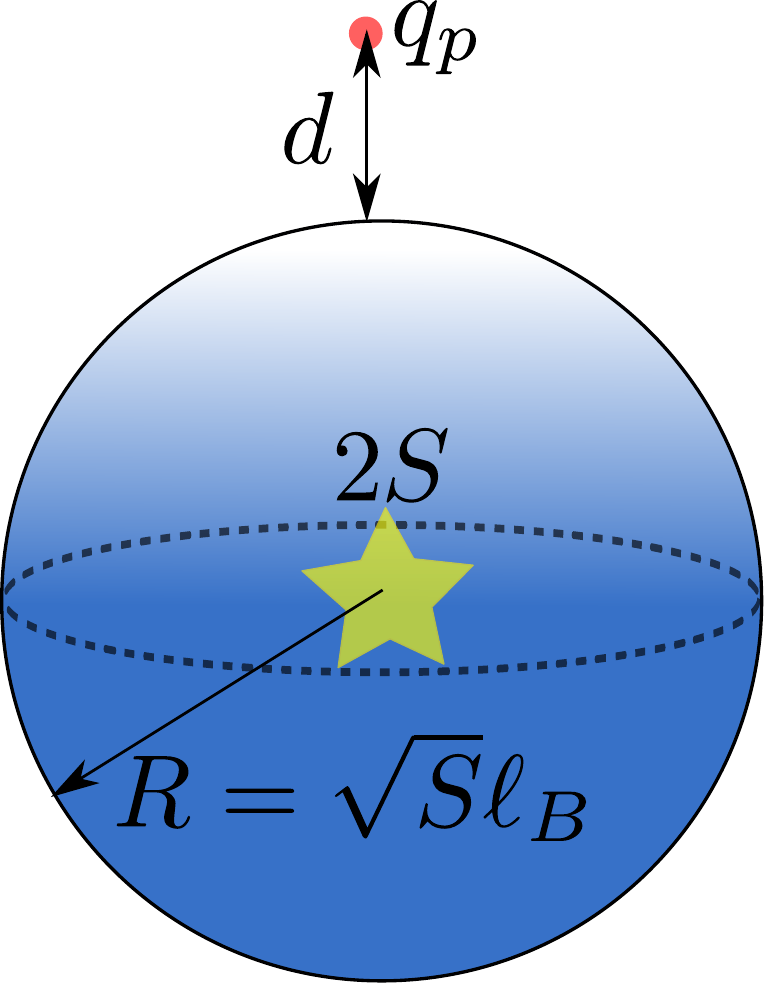}
\vspace{-1mm}
\caption[]{(Color online) Localizing the quasihole on the sphere by an impurity potential above or at the north pole. We consider three types of impurity potentials: (1) projected one-body delta function interaction; (2) Gaussian one-body potential; and (3) Coulomb point charge placed at a distance $d\sim \ell_B$ above the north pole. }
\label{fig:qh}
\vspace{-0pt}
\end{figure}
A practical way to work with Eq.~(\ref{laughlinqh}) is to notice that when $z_0$ corresponds to one of the poles of the sphere, the wave function (\ref{laughlinqh}) still has $L_z$ symmetry and can be expressed as a single Jack polynomial~\cite{jack} labeled by the partition $0100100\ldots1001$ and parameter $\alpha=-2$. This is true only for the pole; wave function for the quasihole localized anywhere else on the sphere is a more complicated superposition of several Jack polynomials~\cite{bernevig}. The same is true for the cylinder as well: a localized quasihole wave function on the cylinder is in general a superposition of several Jack polynomials.

To characterize the spatial extent of the quasihole defect,  we compute the ``excess charge" (integrated deviation of the density from the background density of the ground state)
\begin{equation}
Q(\mathbf{\Omega})=\int_0^{\bf{\Omega}} d^2\mathbf{\Omega}' (\rho_{qh}(\mathbf{\Omega}') - \rho_0).
\end{equation}
Here $\rho_{qh}(\mathbf{\Omega}')$ is the density of the quasihole state at the spherical angle $\mathbf{\Omega}'$, $\rho_0$ is the density of the ground state (a trivial constant on the sphere), and the integral is over the unit sphere ($Q$ as a function of distance $r$ in physical units is obtained by simple rescaling with $\sqrt{S}$ at the end). 

For the quasihole created at $\mathbf{r}=0$ on the cylinder, the formula should be modified as:
\begin{equation}\label{Q_cyl}
Q(x)=\int_{-x}^{x} dx' \int_{0}^{L} dy  (\rho_{qh}(x',y) - \rho_0(x',y)),
\end{equation}
where $L$ is the circumference of the cylinder. The background density $\rho_0$ is no longer a constant due to open boundaries. The charge is integrated over the $y$-axis so that calculating $Q(x)$ requires only the knowledge of diagonal matrix elements of the density operator. It is implicitly assumed that $\rho_{qh}$ and $\rho_0$ have been rescaled to the same size as the number of orbitals in the two cases differs by 1.
 
A criterion for a well-localized quasihole within a certain length $\xi$ is that the limiting value of $Q$ at large distances $r \gtrsim \xi$ approaches $1/3$, but in such a way that it develops a plateau at 1/3 that grows as system size is increased. 

\begin{figure}[htb]
\includegraphics[width=0.7\linewidth,angle=270]{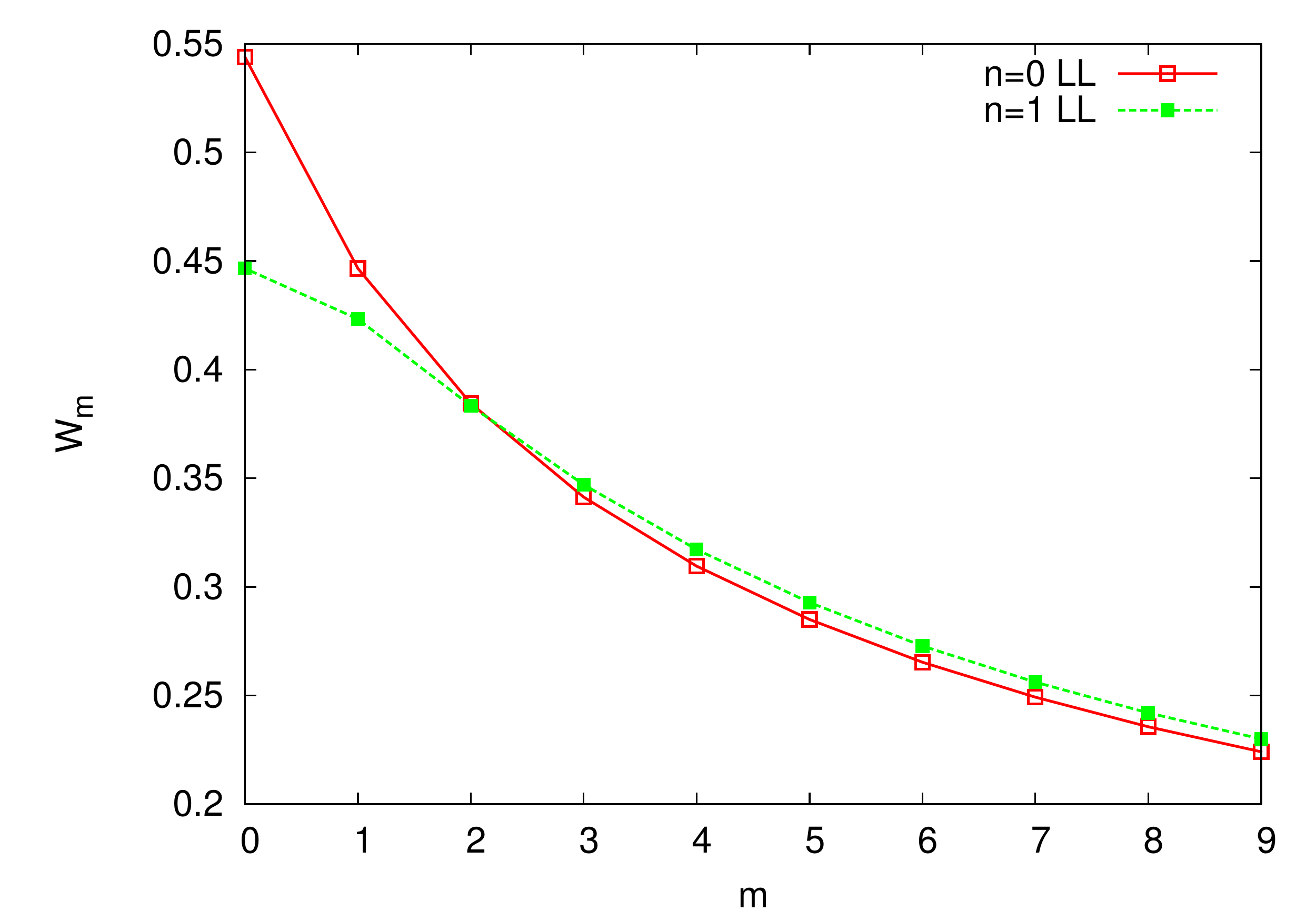}
\vspace{-1mm}
\caption[]{(Color online) Matrix elements $\langle m|V_{imp}|m\rangle$ of the point charge potential, projected into $n=0$ and $n=1$ LL ($q_p=1$, $d=1.4\ell_B$).}
\label{fig:coulombpotential}
\vspace{-0pt}
\end{figure}
Now we discuss different choices for the potential of the impurity probe that is meant to localize the quasihole of the Coulomb interaction ground states at 1/3 or 7/3. As mentioned before, if we have obtained the ground state $\Psi_C$ of the Coulomb Hamiltonian $H_C$, we add a perturbation $V_{imp}(z_0)$ that is not strong enough to cause any major level-crossings in the low-lying part of the spectrum of $H_C$. Then, the combined ground state of $H_C+V_{imp}(z_0)$ would yield a quasihole localized at $z_0$, provided that we have chosen an appropriate form of $V_{imp}$. 

A natural place to look for possible $V_{imp}$ is among short-range one-body potentials such as the projection of delta interaction, Fig.~\ref{fig:qh}:
\begin{equation}\label{deltaimp}
V_{imp}=q_p\delta(z_0)
\end{equation}
Those might, in some sense, model the STM tip brought to the vicinity of two-dimensional electron gas~\cite{toke,morf}. Alternatively, one might consider longer-range potentials with the Gaussian profile parametrized by $\sigma$:
\begin{equation}\label{gaussianimp}
V_{imp}=q_p\exp(-|z-z_0|^2/2\sigma^2).
\end{equation}
It appears to us, however, that the most realistic potential should be the long-range Coulomb potential of a charge $q_p$ brought to some distance $d$ near the electron system~\cite{prodan}. For the sphere, it is convenient to place this charge above the pole, in order to preserve $L_z$ symmetry:
\begin{equation}\label{coulombimp}
V_{imp}=\frac{q_p}{|R\mathbf{\Omega}-(R+d)\hat{z}|}.
\end{equation}
In all the models, $q_p$ is an overall constant that can be varied. Additionally, one can explore further range of potentials by tuning $\sigma$ or $d$ (we have considered the range between $0.1\ell_B$ and $4\ell_B$).       

\begin{figure}[htb]
\includegraphics[width=0.9\linewidth]{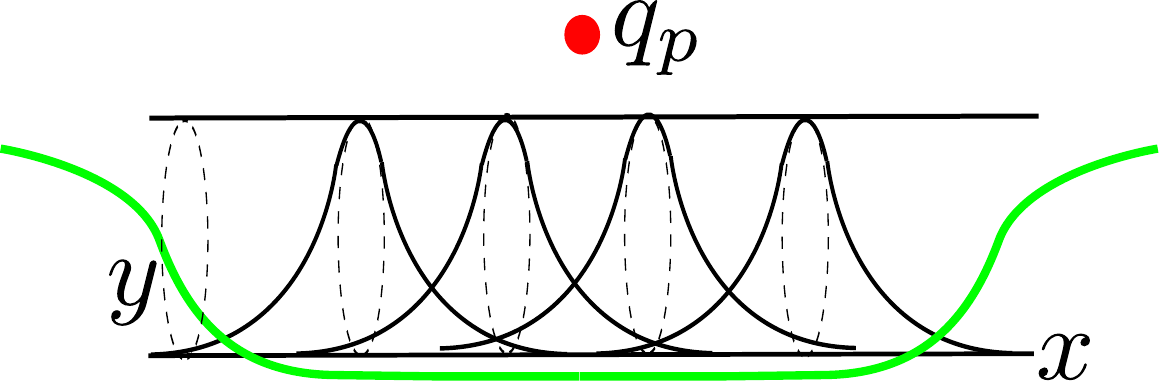}
\vspace{-1mm}
\caption[]{(Color online) Quantum Hall cylinder in a confining potential, with Landau gauge orbitals (periodic along $y$-axis and Gaussian localized along $x$), and an impurity $q_p$. Localizing the quasihole at any given point lifts the symmetry and momentum along the cylinder is no longer a good quantum number.}
\label{fig:cyl}
\vspace{-0pt}
\end{figure}
Having defined the choices for impurity potentials, the actual implementation consists of evaluating the matrix elements $\langle m |V_{imp}|m\rangle$ between the one-body wave functions in the LLL or LL1. Because these wave functions come with different form factors in two LLs, the corresponding matrix elements will also be slightly different, especially in the short-range part, as Fig.~\ref{fig:coulombpotential} illustrates.

We would like to emphasize that cylinder geometry has several features that distinguish it from the sphere. First, the presence of two edges is detrimental for studying the bulk properties of the system, and much larger system sizes are required to reach the bulk of a given size compared the sphere. Secondly, in contrast to other DMRG studies~\cite{zaletel}, we are interested in the long-range Coulomb interaction which requires a confining potential on a finite cylinder. If the state is fragile, like $\nu=7/3$, very fine tuning of the confining potential will be necessary to locate the right phase. Detailed comparison of the confining potentials is explained elsewhere~\cite{sonika}; here we use the potential due to a line of positive charge along the axis of the cylinder, which also ensures the overall charge-neutrality of the system.

Having specified our Hamiltonian, we then represent it in the basis of Slater determinants, $\mathbf{m}\equiv |m_0,m_1,\ldots,m_{2S}\rangle$. Here $m_i$ label the one-particle eigenstates of $L_z$ angular momentum projection on the sphere (or $K_y$ momentum on the cylinder). In the absence of an impurity, total momentum is conserved, and the Hamiltonian can be block-diagonalized in sectors of fixed total $L_z$ or $K_y$. In the presence of an impurity, however, this symmetry will in general be lifted. In the case of the sphere, by positioning the impurity above one of the poles, we are able to preserve the symmetry. This does not hold for the cylinder, where in principle all $K_y$ sectors must be considered simultaneously. In practice, however, we find that it is only necessary to keep a few sectors because of the Gaussian cut-off. For example, if we want to localize a quasihole in the middle of the cylinder ($\mathbf{r}=0$), we keep the sectors $K_y=0,\pm 1, \pm 2, \ldots, \pm \Lambda$, where $\Lambda$ can be varied until convergence is achieved. In order to find the ground state of the Hamiltonian, we use exact diagonalization, as well as density-matrix renormalization (DMRG) group~\cite{feiguin,dlk,zhao} (for details of our DMRG implementation, see Ref.~\onlinecite{dmrg}).   

\section{$\nu=1/3$ and $\nu=7/3$ on the sphere}\label{sphere}

In this section, we present results for the size of the quasihole at $\nu=1/3$ and $\nu=7/3$ in the spherical geometry. Before discussing the quasihole, however, we would like to comment on the nature of the ground state itself in the two cases. While there is little doubt that $\nu=1/3$ Coulomb ground state is described by the Laughlin wave function, there remains some doubt~\cite{jainqh} if the same is true of $\nu=7/3$, because finite systems of up to $N=12$ show very strong finite-size effects. In particular, the counting of entanglement spectrum levels for $\nu=7/3$ matched only a few of the Laughlin state, and the corresponding entanglement gap was small or even non-existent, although it improved when the system size was increased from $N=10$ to $N=12$~\cite{jainqh}.
\begin{figure}[htb]
\includegraphics[width=\linewidth]{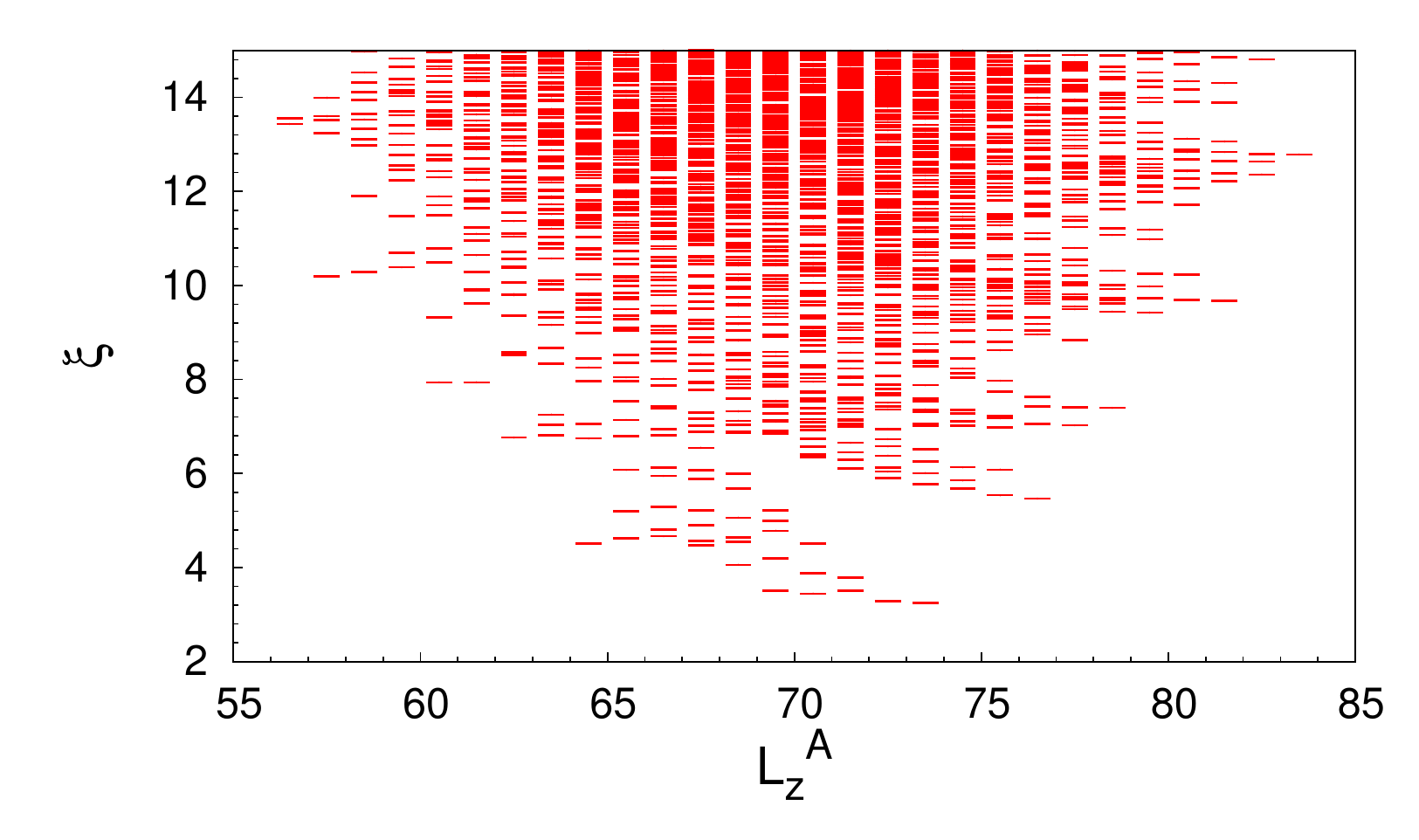}
\vspace{-1mm}
\caption[]{(Color online) Orbital entanglement spectrum of $\nu=7/3$ state on the sphere. Subsystem $A$ contains $l_A=17$ orbitals and $N_A=7$ particles.}
\label{fig:es}
\vspace{-0pt}
\end{figure}

We have extended the calculation in Ref.~\onlinecite{jainqh} by exactly diagonalizing $N=14$ particles at $\nu=7/3$. This calculation was made possible by using an exact Laughlin state (obtained by the Jack generator~\cite{jackgenerator}) as a starting point for the Lanczos algorithm. The corresponding entanglement spectrum is shown in Fig.~\ref{fig:es}, for partitioning the system into two equal subsystems. The momentum of the lowest states corresponds to the one derived from the root partition $100100\ldots1001$, and the counting reads $1,1,2,3,5,6,\ldots$. We observe an entanglement gap separating these levels from the rest of the spectrum, similar to other quantum Hall states, though the value of the gap is small in this case, indicating the fragility of the state. Note that last entanglement level should have a multiplicity of 7, not 6, which we expect to be ``corrected" as the system size is further increased. For the given state, the overlap squared with the Laughlin wave function is 33\%. Given a steady emergence of the Laughlin physics as $N$ is increased from 10 through 14, we believe that the $\nu=7/3$ ground state on the sphere indeed is described by Laughlin's theory.
\begin{figure}[htb]
\includegraphics[angle=270,width=\linewidth]{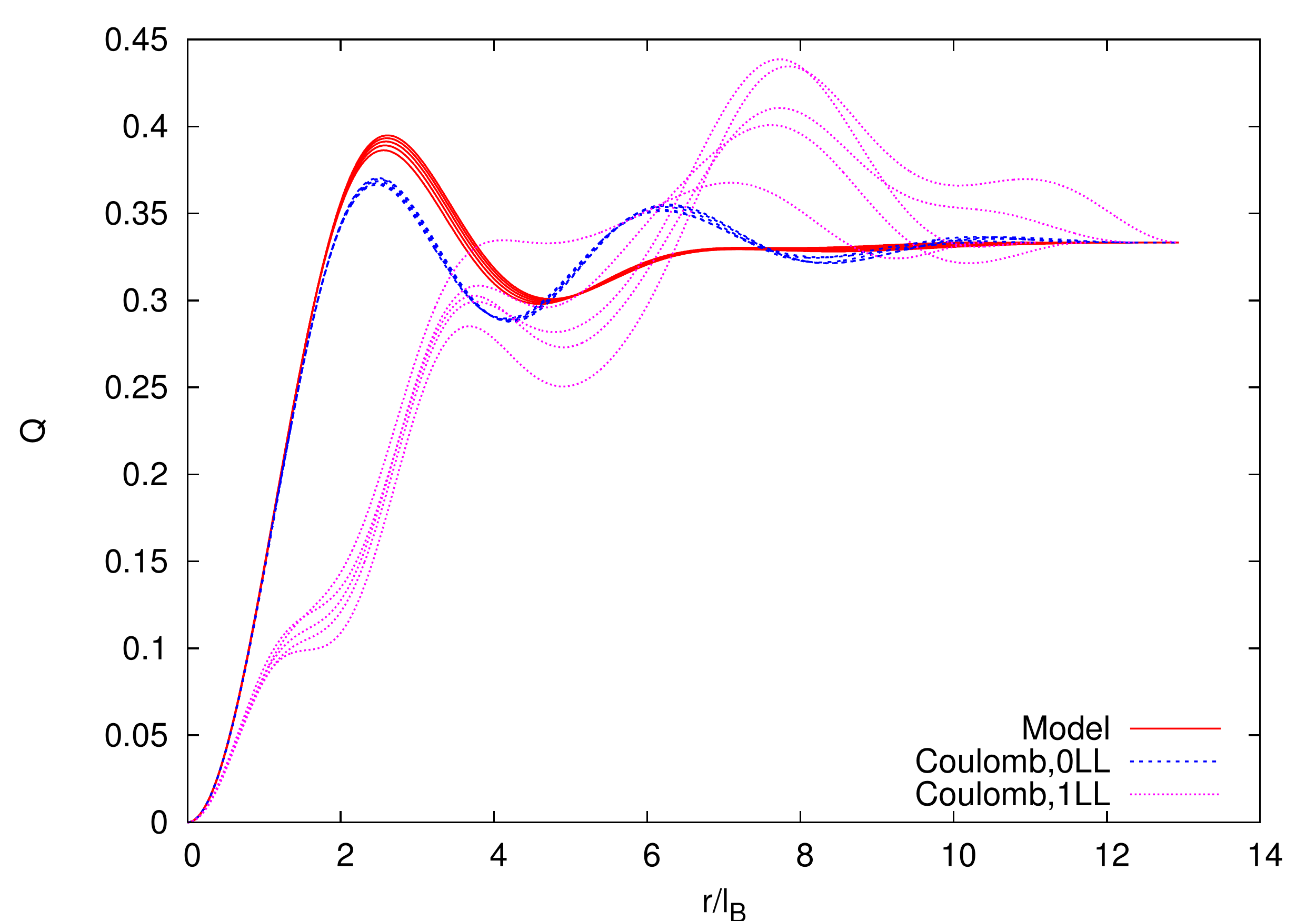}
\vspace{-1mm}
\caption[]{(Color online) Excess charge for the model Laughlin quasihole, and for 1/3 and 7/3 Coulomb ground states with a quasihole pinned using a delta function impurity potential with magnitude $q_p=0.1$. Data is obtained using the Jack polynomials and exact diagonalization for systems $N=8-12$ particles.}
\label{fig:deltaimp}
\vspace{-0pt}
\end{figure}

\begin{figure}[htb]
\includegraphics[angle=270,width=\linewidth]{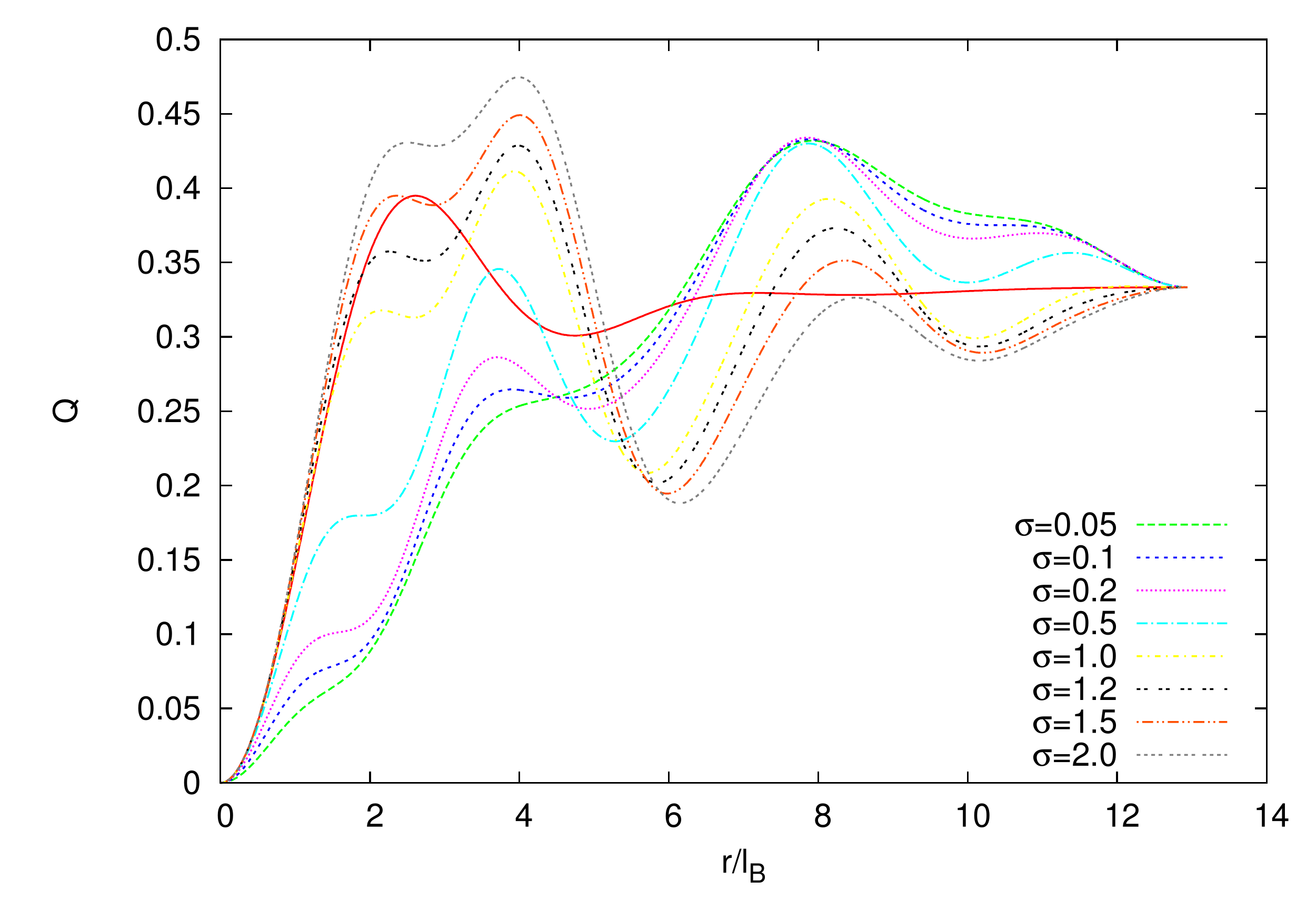}
\vspace{-1mm}
\caption[]{(Color online) Attempt of localizing the $\nu=7/3$ quasihole using the Gaussian potential, Eq.~(\ref{gaussianimp}), for $N=12$ particles. Excess charge is computed for various choices of $\sigma$ and plotted as a function of arc distance on the sphere. For reference, the excess charge for the model Laughlin quasihole is shown in solid line.}
\label{fig:gaussianimp}
\vspace{-0pt}
\end{figure}
Now we proceed to localize a quasihole using a delta function impurity potential, Eq. (\ref{deltaimp}). Data for system sizes $N=8-12$ obtained using Jack polynomials or exact diagonalization is shown in Fig.~\ref{fig:deltaimp}, where we plot the excess charge $Q$ as a function of arc distance on the sphere. We see that the size of the quasihole for the model Laughlin state or the $\nu=1/3$ has fully converged for given systems, and its size can be reliably estimated (see below). On the other hand, for $\nu=7/3$ state, the data varies erratically between different $N$'s and does not converge even for the largest system sizes. This indicates that the delta function impurity potential might be insufficient to localize a quasihole in LL1. 

\begin{figure*}[t]
  \begin{minipage}[l]{\linewidth}
\includegraphics[width=0.9\linewidth]{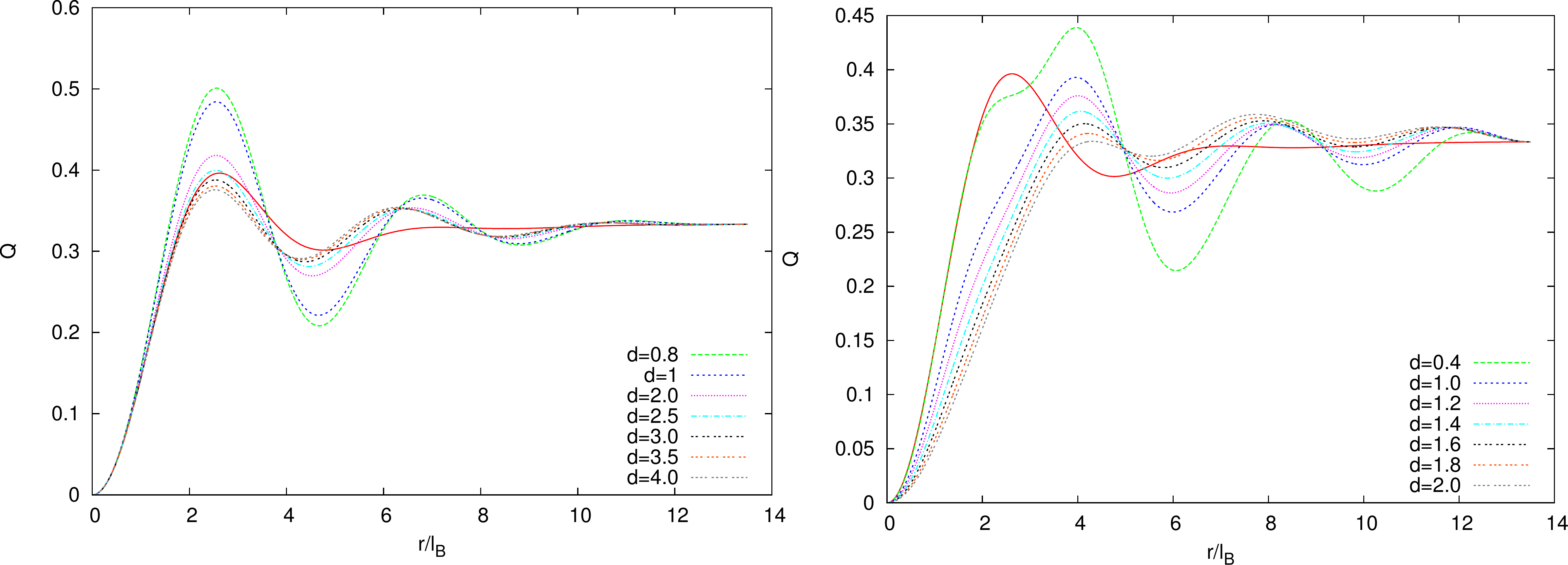}
  \end{minipage}
\caption[]{(Color online) Attempt of localizing the $\nu=1/3$ (left) and $\nu=7/3$ (right) quasihole using the Coulomb point charge potential, Eq.~(\ref{coulombimp}). Excess charge is computed for various choices of distance $d$ of the point charge from the north pole (Fig.~\ref{fig:qh}), and plotted as a function of arc distance on the sphere. For reference, the excess charge for the model Laughlin quasihole is shown in solid line. Data is obtained by exact diagonalization for $N=13$ particles.}
\label{fig:coulombimp}
\vspace{-0pt}
\end{figure*}          
To explore whether a longer range potential, such as the Gaussian potential in Eq.~(\ref{gaussianimp}), might work better in localizing the quasihole, in Fig.~\ref{fig:gaussianimp} we compute the excess charge as a function of the characteristic length scale $\sigma$ of the Gaussian potential. For the given system size ($N=12$), we find the Gaussian potential is also unable to localize the quasihole. Thus, the problem of finding the right pinning potential is subtle.

Finally, we present the results for the quasihole localized using a point charge situated at a distance $d$ from the pole of the sphere, Fig.~\ref{fig:coulombimp}. The charge generates a long-range pinning potential defined by Eq.~(\ref{coulombimp}), which can be tuned through its magnitude $q_p$ and distance $d$. In Fig.~\ref{fig:coulombimp}, we fix $q_p=1$ and vary $d$, which enables us to localize the $\nu=7/3$ quasihole. For comparison, we also show the data for $\nu=1/3$ quasihole localized using the same potential. If $d$ is not too small, $\nu=1/3$ charge profile is very similar to the one obtained using a delta function impurity potential, Fig.~\ref{fig:deltaimp}. If $d$ becomes much smaller than $\ell_B$, the oscillations of the charge become very pronounced. Through some trial and error, we have established the following range of $d$ as optimal:
\begin{equation}\label{optimald}
\nu=1/3:  \; d^* \approx 3\ell_B; \; \; \; \nu=7/3: \; d^* \approx 1.3-1.4\ell_B.
\end{equation}

\begin{figure}[htb]
\includegraphics[angle=270,width=\linewidth]{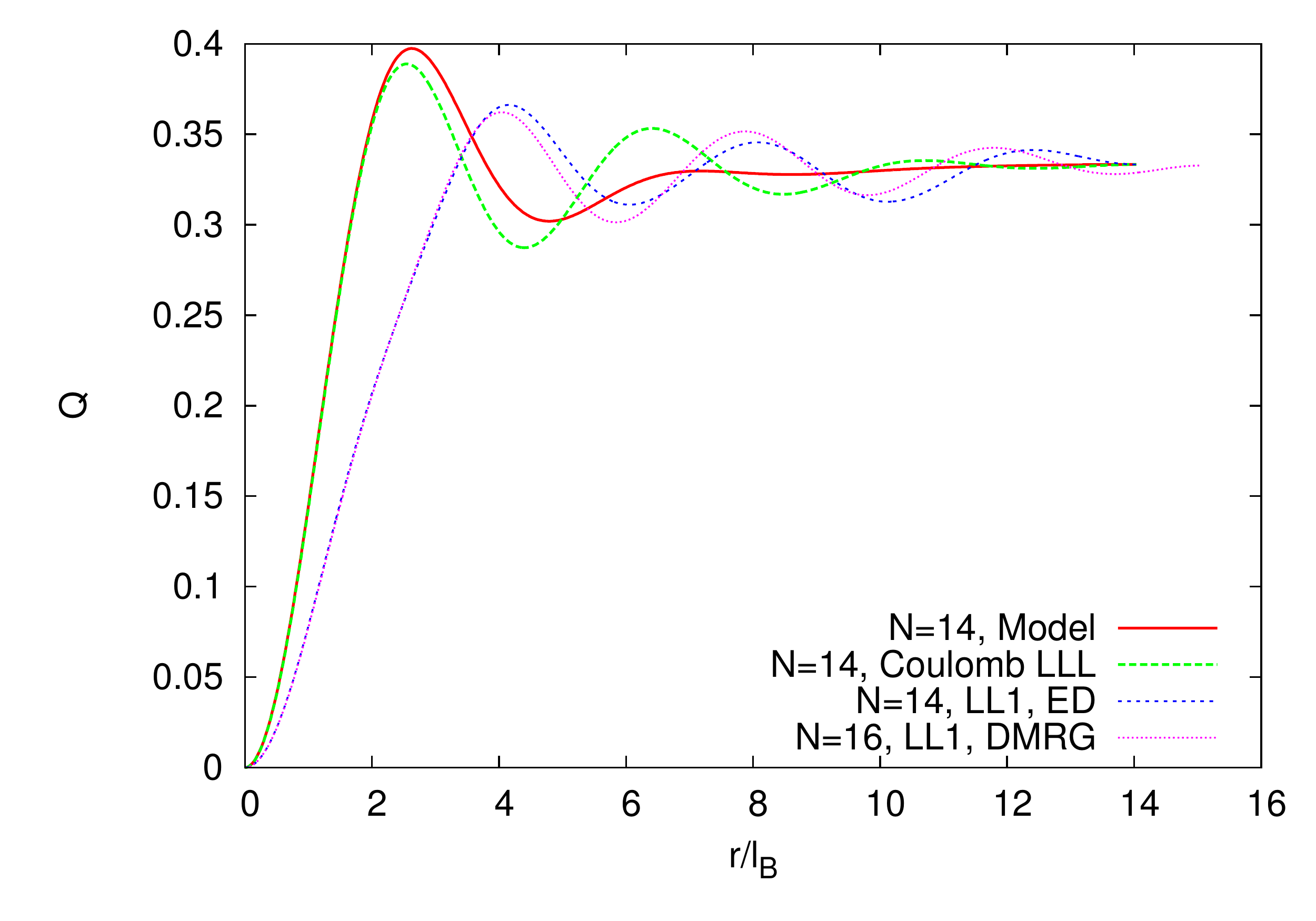}
\vspace{-1mm}
\caption[]{(Color online) Summary of our best results for localizing the quasihole on the sphere: model quasihole (obtained using Jack polynomials), LLL Coulomb quasihole ($N=14$, localized using point charge at distance $d=3\ell_B$), and LL1 Coulomb quasihole (for $N=14$, obtained using exact diagonalization, and $N=16$ using DMRG (number of states kept is $8000$, discarded entropy is $10^{-12}$), with $d=1.4\ell_B$).}
\label{fig:spherebest}
\vspace{-0pt}
\end{figure}

Using these optimal values $d^*$, in Fig.~\ref{fig:spherebest} we summarize our main results for the sphere geometry: $N=14$ particles using exact diagonalization, and $N=16$ via DMRG. Unfortunately, as established previously, spherical geometry is not suitable for DMRG because of curvature effects, and our data is limited to $N=16$, though even larger systems would be desirable. However, the main part of the curve (until the first maximum, $\lesssim 4\ell_B$) is fully converged, and it is this part that dominates the moments of the density distribution that define the size of the quasihole (see below). Moreover, further oscillations originating from the long-range tail show a clear decay to the background density, supporting the fact that the defect has been successfully localized.    

In the remainder of this section, we analyze the data shown in Fig.~\ref{fig:spherebest} using various criteria to obtain estimates of the quasihole size. First of all, by direct visual inspection, we see 
that the curves for $Q$ of the model Laughlin state and $1/3$ Coulomb state almost overlap until the first peak at $r\approx 2.5\ell_B$. Beyond this point, the model quasihole completely relaxes to the background around $r\approx 6\ell_B$, while the Coulomb quasihole has additional smaller oscillations around the mean $1/3$ value. These oscillations eventually die out by $r\approx 10\ell_B$. For $\nu=7/3$, the first peak is farther out ($r\approx 4\ell_B$), and additional peaks seem to fully disappear only in our largest system ($N=16$). Therefore, our upper estimates for radii of the quasiholes are:
\begin{equation}
R_{Laughlin}\approx 6\ell_B; \; \; R_{1/3}\approx 10\ell_B; \; \; R_{7/3}\gtrsim 15\ell_B.
\end{equation}  
However, because the distribution of charge is such that it contains a dominant first peak and smaller oscillatory decay superposed on it, we believe the above values present an overestimate of the quasihole size. 

One way to quantify the extent of the quasihole is through the moments of the density distribution. The first moment is defined as:
\begin{equation}
R_1=\sqrt{S} \frac{\int_{0}^{\pi}|\rho(\theta)-\rho(\pi)|\theta \sin\theta d\theta}{\int_{0}^{\pi} |\rho(\theta)-\rho(\pi)| \sin\theta d\theta}.
\end{equation}
The square root of the second moment gives the standard deviation:
\begin{equation}
R_2=\sqrt{S}\sqrt{\frac{\int_{0}^{\pi}|\rho(\theta)-\rho(\pi)|\theta^2 \sin\theta d\theta}{\int_{0}^{\pi} |\rho(\theta)-\rho(\pi)| \sin\theta d\theta}}
\end{equation}
Table 1 gives the values of $R_1$ and $R_2$ for the results in Fig.~\ref{fig:spherebest}.
\begin{table}
\centering
    \begin{tabular}{| l | l | l | l |}
    \hline
    Interaction   & N  & $R_1$  & $R_2$  \\ \hline
    Model Laughlin          & 14 & 2.0 & 2.5  \\ \hline
    Coulomb, LL=0 & 14 & 3.1 & 4.2  \\ \hline
    Coulomb, LL=1 & 14 & 5.0 & 6.1 \\ \hline
    Coulomb, LL=1 & 16 & 5.7 & 6.9  \\
    \hline
    \end{tabular}
\caption{Size of the quasihole evaluated through moments of the density distribution.}
\end{table}

We can also fit the maxima of $Q$ using an exponentially decaying curve:
\begin{equation}
|Q(\theta)-\frac{1}{3}|=\exp(a\theta+b)
 \end{equation}
 The length scale of decay then is $R_{fit}=\sqrt{S}/a$, shown in Table II.

\begin{table}
    \begin{tabular}{| l | l | l | l |}
    \hline
    Interaction   & N  &  $R_{fit}$  \\ \hline
    Model Laughlin          & 14 &  2.4   \\ \hline
    Coulomb, LL=0 & 14 &  4.1 \\ \hline
    Coulomb, LL=1 & 14 &  7.0   \\ \hline
    Coulomb, LL=1 & 16 &  7.0 \\
    \hline
    \end{tabular}
\caption{Estimated radius of the quasihole from the exponential decay of $Q$ in Fig.~\ref{fig:spherebest}.}
\end{table}

\begin{figure}[htb]
\includegraphics[,width=0.9\linewidth]{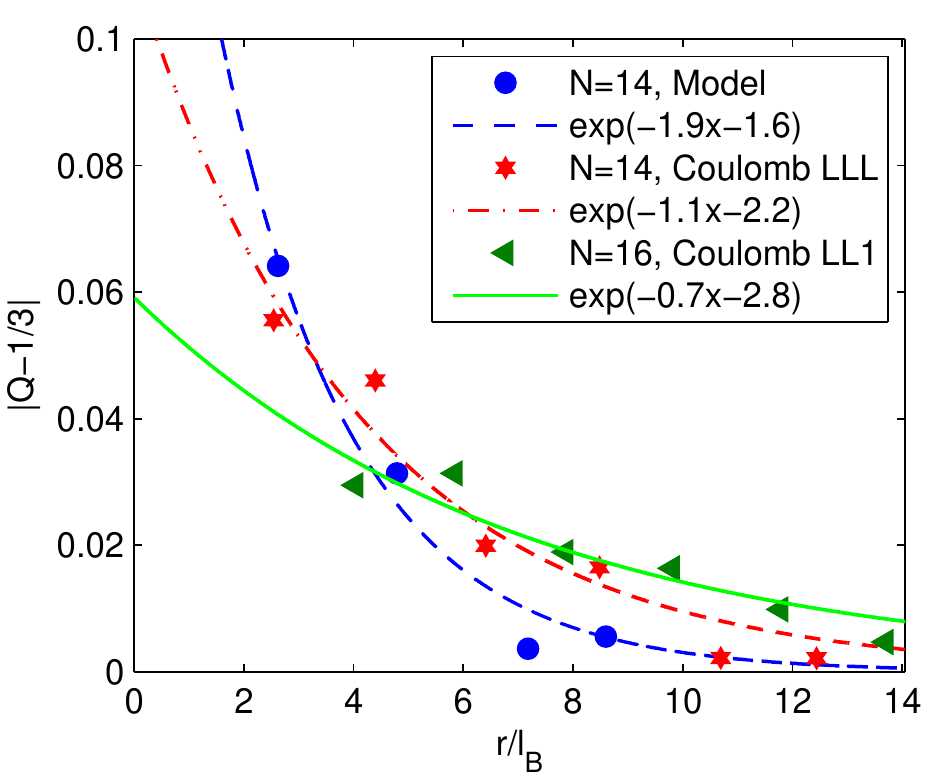}
\vspace{-1mm}
\caption[]{(Color online) Fits of the maxima of the accumulated charge $|Q(\theta)-\frac{1}{3}|$ for the  results shown in Fig.~\ref{fig:spherebest}.}
\label{fig:fit_extremes}
\vspace{-0pt}
\end{figure}

Reconciling the quantitative analysis above, we arrive at the following estimates for the size of the quasihole on the sphere:
\begin{equation}\label{sizebest}
R_{Laughlin}\approx 2.5\ell_B; \; \; R_{1/3}\approx 4\ell_B; \; \; R_{7/3}\gtrsim 7\ell_B.
\end{equation}  
Therefore, our main conclusion (in qualitative agreement with Ref.~\onlinecite{jainqh}) is that the $7/3$ quasihole is bigger in size from the $1/3$ quasihole by a factor of 2.  

\section{$\nu=1/3$ and $\nu=7/3$ on the cylinder}\label{cylinder}

In this section we repeat the previous calculation for the cylinder geometry. The motivation for this is two-fold. Firstly, independent results obtained with a different boundary condition would represent a good confirmation of the validity of our results on the sphere. More importantly, it has been shown that cylinder geometry is particularly suitable for DMRG, and thus, as we will see below, we will be able to access larger system sizes than on the sphere. 

In the cylinder geometry, Fig.~\ref{fig:cyl} we consider periodic boundary condition along $y$-axis and open edges along $x$-direction. The quasihole is created in the middle of the cylinder. Since we are interested in localizing the quasihole at a point, it is not possible to preserve the translational symmetry along the perimeter, and obtaining even the localized Laughlin quasihole becomes a non-trivial problem. 

\begin{figure}[htb]
\includegraphics[width=0.9\linewidth]{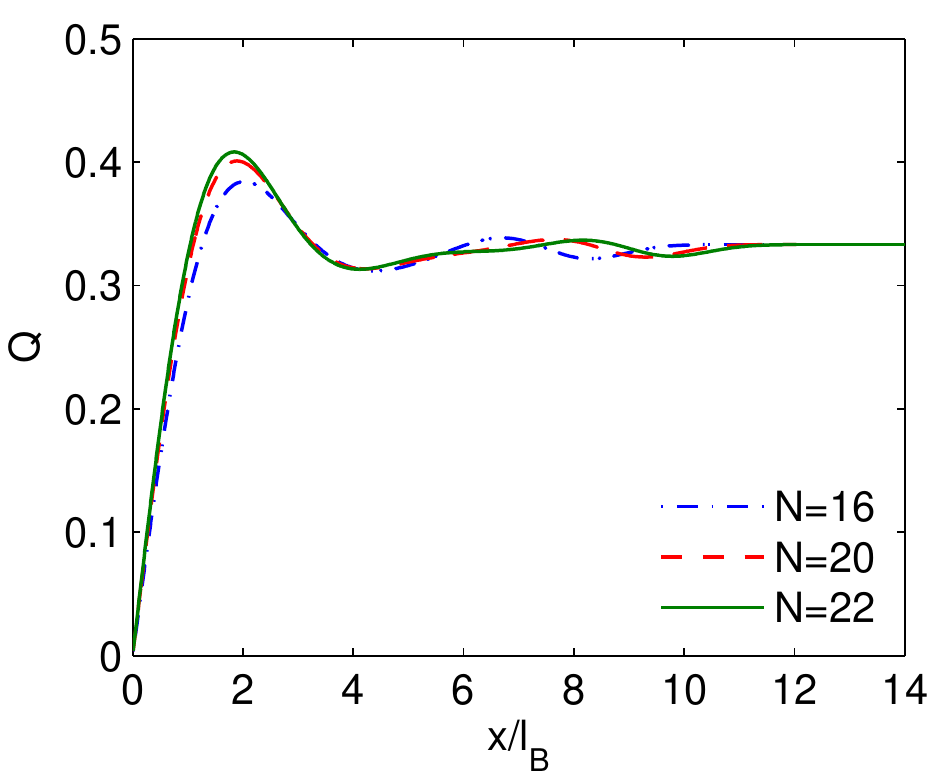}
\vspace{-1mm}
\caption[]{(Color online) Integrated excess charge for the model Laughlin quasihole pinned using a delta function impurity potential with magnitude $q_p=1$ on the cylinder with aspect ratio 1. Data is obtained using DMRG (number of states kept is $1500$, discarded entropy is $10^{-12}$) in momentum sectors $K_y=0,\ldots, \pm 3$.}
\label{fig:qh_delta}
\vspace{-0pt}
\end{figure}
As explained in Sec.~\ref{method}, on a cylinder we must expand the Hilbert space to include several $K_y$ momentum sectors. To obtain the model quasihole, we diagonalize the $V_1$ Haldane pseudopotential with a delta function impurity potential (Eq.~(\ref{deltaimp})). We plot the integrated excess charge (Eq.~(\ref{Q_cyl})) $Q(x)$, evaluated from the center of the cylinder ($x=0$) towards the edges, in Fig.~\ref{fig:qh_delta}. Note that the curves look very similar to the sphere data, except for an additional peak that moves in the interval between $6\ell_B$ and $8\ell_B$. This peak is moving outward as we increase $N$, and represents an edge effect. Unfortunately, as we see from this data, the presence of the edge strongly affects the bulk properties, and we need essentially twice as many particles on the cylinder ($N\approx 20$) to obtain a similar quality of data as for the sphere with $N\sim 10$.       

\begin{figure}[htb]
\includegraphics[,width=0.9\linewidth]{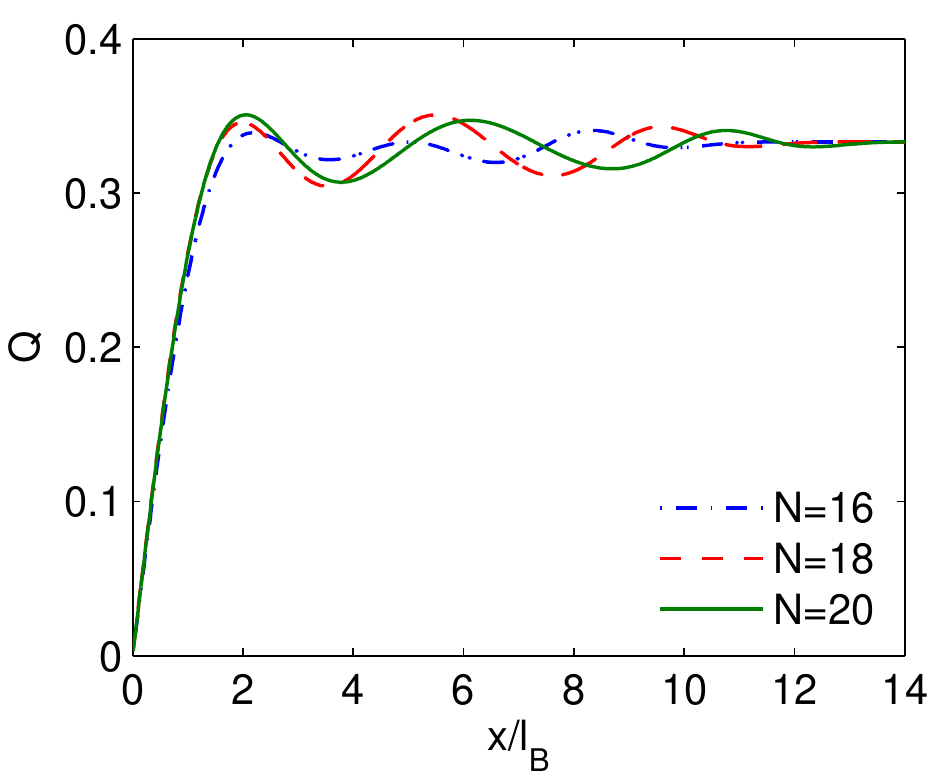}
\vspace{-1mm}
\caption[]{(Color online) Integrated excess charge for the quasihole at filling factor $1/3$, pinned using a delta function impurity potential with magnitude $q_p=1$ on the cylinder with circumference $L=15 \ell_B$. Data is obtained using DMRG (number of states kept is $2500$, discarded entropy is $10^{-12}$) in momentum sectors $K_y=0,\ldots,\pm 3$.}
\label{fig:qh_1third}
\vspace{-0pt}
\end{figure}
\begin{table}[h!]
    \begin{tabular}{| l | l | l | l | l | l | l | l | l | l |}
    \hline
    Geometry & Type & N &  1 & 2 & 3 & 4 & 5 & 6  & 7 \\ \hline
    Cylinder &  Model & 22 &  0 &   1.8 &    4.1 &    8.2 &    9.8 &   &  \\ \hline
    Cylinder & 1/3 & 20 &  0  &  2.1 &   3.8 &    6.1 &    8.7 &   10.8 &  12.3   \\ \hline
    Sphere & Model & 14 &  0  &  2.6 &    4.8 &    7.2 &    8.6 & & \\ \hline
    Sphere & 1/3 & 14 &   0  & 2.5 &    4.4 &    6.4 &     8.5 &   10.7 &   12.4 \\ \hline
    Sphere & 7/3 & 14 & 0  & 4.2 &  6.1 &    8.1 &   10.1 &  12.4 & \\ \hline
    Sphere & 7/3 & 16 &  0 & 4.0 &  5.9 &    7.9 &   9.8 &  11.8 &   13.7  \\ \hline
    \end{tabular}
\caption{Comparison of the values of $r/l_B$ or $x/l_B$ at the extrema of $Q$, for different sizes $N$ for the sphere and cylinder.}
\end{table}

In Fig.~\ref{fig:qh_1third} we show the same calculation for $1/3$ Coulomb quasihole, for several system sizes $N$ and fixed perimeter $L=15\ell_B$. This data illustrates that we have correctly captured the location of the first peak, but the second peak is still developing and moving outward as a function of $N$. As it moves outward, the magnitude of this peak is expected to drop, and the curve should increasingly resemble the sphere data. In Table III we show a detailed comparison of the extrema of $Q$ on the cylinder and the sphere. We see that the second peak is located at a larger distance on the sphere, which is consistent with our previous claim that the second peak has not fully converged by $N=20$, and is still expected to move outwards with increasing $N$. 

Unfortunately, for $\nu = 7/3$ on the cylinder, we have not been able to find a robust incompressible ground state for the bare Coulomb interaction (i.e, excluding finite width, Landau-level mixing, modification of pseudopotentials etc.) in finite systems of up to $N=22$ particles on a cylinder. We define our target state as the one that has
constant density in the bulk, and the counting of the entanglement spectrum that matches the Laughlin state, for at least several low-lying momentum sectors. Although we have used several types of confining potentials (parabolic, line charge etc.), none of them has consistently produced a ground state that fulfills both of these criteria. Additionally, the overlap with the Laughlin wavefunction was found to decrease rapidly as $N$ was increased from 8 to 12. The reason for such a weak $7/3$ state could be the aggravated edge effect in this case that propagates further into the bulk due to the very fragile entanglement gap, Fig.~\ref{fig:es}.

\section{Conclusions}\label{conclusions}

In this work we have addressed a basic question: what is the size of the quasiholes in $\nu=1/3$ and $\nu=7/3$ quantum Hall states. We have performed a systematic study using a variety of pinning potentials that localize the quasihole, in both spherical and cylindrical geometry, employing exact diagonalization as well as DMRG. With these unbiased calculations, we have obtained reliable bounds on the size of the quasiholes in the two cases, which are given in Eq.~(\ref{sizebest}), with the corresponding charge profiles shown in Fig.~\ref{fig:spherebest}. These results have direct implications for any type of experiment that involves transport or braiding of quasiparticles. We emphasize, however, that our calculation assumes zero thickness and does not include the Landau-level mixing effects. These effects might slighly renormalize our results in real samples. 

For simplicity, in this work we have focused entirely on the quasihole excitations, as opposed to the quasielectron. We expect that the size of the quasielectron should be roughly similar to that of the quasihole, but not necessarily the same. This issue, along with the sizes of more complicated (especially non-Abelian) states remains to be addressed in future work. Another open question resulting from current work is the nature of the $\nu=7/3$ state on the cylinder. If the reason for its absence can be attributed to the edge effect, then DMRG implementations on an infinite cylinder~\cite{zaletel} should be able to find an incompressible state. However, thus far such implementations have been limited to artificial short-range interactions, whereas our study applies to a realistic system described by the long-range Coulomb interaction.

{\sl Acknowledgements}. This work was supported by DOE grant DE-SC$0002140$. R. N. B. acknowledges the hospitality of the Institute for Advanced Study, Princeton while this paper was being written.


\begin{thebibliography}{99}

\bibitem{laughlin}
R. B. Laughlin, Phys. Rev. Lett. {\bf 50}, 1395 (1983).

\bibitem{rh85}
F. D. M. Haldane and E. H. Rezayi, Phys. Rev. Lett. {\bf 54}, 237  (1985).

\bibitem{tsg}
D. C. Tsui, H. L. Stormer, and A. C. Gossard, Phys. Rev. Lett. {\bf 48}, 1559 (1982).

\bibitem{kumar}
A. Kumar, G. A. Cs\'athy, M. J. Manfra, L. N. Pfeiffer, and K. W. West, 
Phys. Rev. Lett. {\bf 105}, 246808 (2010). 

\bibitem{prange}
\emph{The Quantum Hall Effect}, 2nd ed., edited by R. E. Prange and S. M. Girvin, Springer-Verlag, New York, 1990.

\bibitem{Moore91}
G. Moore and N. Read, Nucl. Phys. B {\bf 360}, 362 (1991).

\bibitem{white}
S. White, Phys. Rev. Lett. {\bf 69}, 2863 (1992).

\bibitem{prds}
Z. Papi\'c, N. Regnault, and S. Das Sarma, Phys. Rev. B {\bf 80}, 201303 (2009); M.R. Peterson, Th. Jolicoeur, and S. Das Sarma, Phys. Rev. B {\bf 78}, 155308 (2008). 

\bibitem{jainqh}
Ajit C. Balram, Ying-Hai Wu, G. J. Sreejith, Arkadiusz W\'ojs, and Jainendra K. Jain,
Phys. Rev. Lett. {\bf 110}, 186801 (2013).

\bibitem{jack} B. Andrei Bernevig and F. D. M. Haldane, Phys. Rev. Lett. {\bf 100}, 246802 (2008).

\bibitem{mps}
Michael P. Zaletel and Roger S. K. Mong, Phys. Rev. B {\bf 86}, 245305 (2012); 
B. Estienne, Z. Papi\'c, N. Regnault, and B. A. Bernevig, Phys. Rev. B {\bf 87}, 161112 (2013). 

\bibitem{prodan}
Emil Prodan and F. D. M. Haldane, Phys. Rev. B {\bf 80}, 115121 (2009).

\bibitem{toke}
Csaba T\H{o}ke, Nicolas Regnault, and Jainendra K. Jain, Phys. Rev. Lett. {\bf 98}, 036806 (2007).   

\bibitem{morf}
M. Storni and R. H. Morf, Phys. Rev. B {\bf 83}, 195306 (2011).

\bibitem{willett}R. L. Willett, L. N. Pfeiffer, and K. W. West, Phys. Rev. B 82, 205301 (2010)

\bibitem{camino}F. E. Camino, W. Zhou, and V. J. Goldman, Phys. Rev. B 72, 075342 (2005)

\bibitem{Haldane86}
F. D. M. Haldane, Phys. Rev. Lett. {\bf 51}, 605 (1983).

\bibitem{hr_cylinder} E. H. Rezayi and F. D. M. Haldane, Phys. Rev. B {\bf 50}, 17199 (1994). 

\bibitem{jainbook} J. K. Jain, \emph{Composite fermions}, (Cambridge University Press, 2007).

\bibitem{sonika} S. Johri \emph{et al.}, (unpublished).

\bibitem{arovas} Daniel Arovas, J. R. Schrieffer, and Frank Wilczek, Phys. Rev. Lett. {\bf 53}, 722 (1984). 

\bibitem{bernevig} B. A. Bernevig and F. D. M. Haldane, Phys. Rev. Lett. {\bf 102}, 066802 (2009)

\bibitem{zaletel} Michael P. Zaletel, Roger S. K. Mong, and Frank Pollmann, Phys. Rev. Lett. {\bf 110}, 236801 (2013).

\bibitem{feiguin} A. E. Feiguin, E. Rezayi, C. Nayak, and S. Das Sarma, Phys. Rev. Lett. {\bf 100}, 166803 (2008).

\bibitem{dlk} D. L. Kovrizhin, Phys. Rev. B {\bf 81}, 125130 (2010).

\bibitem{zhao} Jize Zhao, D. N. Sheng, and F. D. M. Haldane, Phys. Rev. B {\bf 83}, 195135 (2011).

\bibitem{dmrg} Zi-Xiang Hu, Z. Papic, S. Johri, R. N. Bhatt, and Peter Schmitteckert, Phys. Lett. A {\bf 376}, 2157 (2012). 

\bibitem{jackgenerator} http://www.nick-ux.org/~regnault/jack/‎

\end{thebibliography}
\end{document}